\documentclass[pdftex,epjc3upd,twocolumn]{svjour3}

\RequirePackage{graphicx}
\usepackage{tikz}
\RequirePackage{mathptmx}
\usepackage[english]{babel}
\usepackage{amsmath}
\usepackage{mathtools}
\usepackage{siunitx}
\usepackage{comment}
\usepackage{tabularx}
\usepackage[version=4]{mhchem}
\usepackage[font={small},labelfont=bf,labelsep=quad]{caption}
\usepackage[font={normal}]{subcaption}
\usepackage{cite}
\usepackage{upgreek}
\RequirePackage[colorlinks,citecolor=blue,urlcolor=blue,linkcolor=blue]{hyperref}

\defineshorthand{"~}{\babelhyphen{nobreak}}
\useshorthands{"}

\usepackage[switch]{lineno}

\newcommand{\geant}{Geant4}

\makeatletter
\@ifundefined{@setmarks}{\let\@setmarks\relax}{}
\makeatother

\begin{document}
\sloppy

\title{Cosmogenic background simulations for the DARWIN observatory at different underground locations}

\date{Received: date / Accepted: date}

\author{M.~Adrover\thanksref{addr17}
\and
L.~Althueser\thanksref{addr7}
\and
B.~Andrieu\thanksref{addr18}
\and
E.~Angelino\thanksref{addr14}
\and
J.~R.~Angevaare\thanksref{addr8}
\and
B.~Antunovic\thanksref{addr33,addr46}
\and
E.~Aprile\thanksref{addr3}
\and
M.~Babicz\thanksref{addr17}
\and
D.~Bajpai\thanksref{addr38}
\and
E.~Barberio\thanksref{addr34}
\and
L.~Baudis\thanksref{addr17}
\and
M.~Bazyk\thanksref{addr13}
\and
N.~Bell\thanksref{addr34}
\and
L.~Bellagamba\thanksref{addr0}
\and
R.~Biondi\thanksref{addr6}
\and
Y.~Biondi\thanksref{addr26}
\and
A.~Bismark\thanksref{addr17}
\and
C.~Boehm\thanksref{addr35}
\and
A.~Breskin\thanksref{addr16}
\and
E.~J.~Brookes\thanksref{addr8}
\and
A.~Brown\thanksref{addr19}
\and
G.~Bruno\thanksref{addr9}
\and
R.~Budnik\thanksref{addr16}
\and
C.~Capelli\thanksref{addr17}
\and
J.~M.~R.~Cardoso\thanksref{addr2}
\and
A.~Chauvin\thanksref{addr31}
\and
A.~P.~Cimental~Chavez\thanksref{addr17}
\and
A.~P.~Colijn\thanksref{addr8}
\and
J.~Conrad\thanksref{addr12}
\and
J.~J.~Cuenca-Garc\'ia\thanksref{addr17,addr26}
% ...
(DARWIN Collaboration\thanksref{email1}). }
\newcommand{\bologna}{Department of Physics and Astronomy, University of Bologna and INFN-Bologna, 40126 Bologna, Italy}
\newcommand{\chicago}{Department of Physics \& Kavli Institute for Cosmological Physics, University of Chicago, Chicago, IL 60637, USA}
\newcommand{\coimbra}{LIBPhys, Department of Physics, University of Coimbra, 3004-516 Coimbra, Portugal}
\newcommand{\columbia}{Physics Department, Columbia University, New York, NY 10027, USA}
\newcommand{\lngs}{INFN-Laboratori Nazionali del Gran Sasso and Gran Sasso Science Institute, 67100 L'Aquila, Italy}
\newcommand{\mainz}{Institut f\"ur Physik \& Exzellenzcluster PRISMA$^{+}$, Johannes Gutenberg-Universit\"at Mainz, 55099 Mainz, Germany}
\newcommand{\mpik}{Max-Planck-Institut f\"ur Kernphysik, 69117 Heidelberg, Germany}
\newcommand{\munster}{Institut f\"ur Kernphysik, Westf\"alische Wilhelms-Universit\"at M\"unster, 48149 M\"unster, Germany}
\newcommand{\nikhef}{Nikhef and the University of Amsterdam, Science Park, 1098XG Amsterdam, Netherlands}
\newcommand{\nyuad}{New York University Abu Dhabi - Center for Astro, Particle and Planetary Physics, Abu Dhabi, United Arab Emirates}
\newcommand{\purdue}{Department of Physics and Astronomy, Purdue University, West Lafayette, IN 47907, USA}
\newcommand{\rice}{Department of Physics and Astronomy, Rice University, Houston, TX 77005, USA}
\newcommand{\stockholm}{Oskar Klein Centre, Department of Physics, Stockholm University, AlbaNova, Stockholm SE-10691, Sweden}
\newcommand{\subatech}{SUBATECH, IMT Atlantique, CNRS/IN2P3, Universit\'e de Nantes, Nantes 44307, France}
\newcommand{\torino}{INAF-Astrophysical Observatory of Torino, Department of Physics, University  of  Torino and  INFN-Torino,  10125  Torino,  Italy}
\newcommand{\ucsd}{Department of Physics, University of California San Diego, La Jolla, CA 92093, USA}
\newcommand{\wis}{Department of Particle Physics and Astrophysics, Weizmann Institute of Science, Rehovot 7610001, Israel}
\newcommand{\zurich}{Physik-Institut, University of Z\"urich, 8057  Z\"urich, Switzerland}
\newcommand{\lpnhe}{LPNHE, Sorbonne Universit\'{e}, CNRS/IN2P3, 75005 Paris, France}
\newcommand{\freiburg}{Physikalisches Institut, Universit\"at Freiburg, 79104 Freiburg, Germany}
\newcommand{\napels}{Department of Physics ``Ettore Pancini'', University of Napoli and INFN-Napoli, 80126 Napoli, Italy}
\newcommand{\nagoya}{Kobayashi-Maskawa Institute for the Origin of Particles and the Universe, and Institute for Space-Earth Environmental Research, Nagoya University, Furo-cho, Chikusa-ku, Nagoya, Aichi 464-8602, Japan}
\newcommand{\laquila}{Department of Physics and Chemistry, University of L'Aquila, 67100 L'Aquila, Italy}
\newcommand{\tokyo}{Kavli Institute for the Physics and Mathematics of the Universe (WPI), The University of Tokyo, Higashi-Mozumi, Kamioka, Hida, Gifu 506-1205, Japan}
\newcommand{\kobe}{Department of Physics, Kobe University, Kobe, Hyogo 657-8501, Japan}
\newcommand{\ucla}{Physics \& Astronomy Department, University of California, Los Angeles, CA 90095, USA}
\newcommand{\kitiap}{Institute for Astroparticle Physics, Karlsruhe Institute of Technology, 76021 Karlsruhe, Germany}
\newcommand{\kitetp}{Institute of Experimental Particle Physics, Karlsruhe Institute of Technology, 76021 Karlsruhe, Germany}
\newcommand{\kitipe}{Institute for Data Processing and Electronics, Karlsruhe Institute of Technology, 76021 Karlsruhe, Germany}
\newcommand{\tsinghua}{Department of Physics \& Center for High Energy Physics, Tsinghua University, Beijing 100084, China}
\newcommand{\ferrara}{INFN - Ferrara and Dip. di Fisica e Scienze della Terra, Universit\`a di Ferrara, 44122 Ferrara, Italy}
\newcommand{\heidelberg}{Physikalisches Institut, Universit\"at Heidelberg, Heidelberg, Germany}
\newcommand{\heidelbergki}{Kirchhoff-Institut f\"ur Physik, Universit\"at Heidelberg, Heidelberg, Germany}
\newcommand{\belgrade}{Vinca Institute of Nuclear Science, University of Belgrade, Mihajla Petrovica Alasa 12-14. Belgrade, Serbia}
\newcommand{\melbourne}{ARC Centre of Excellence for Dark Matter Particle Physics, School of Physics, The University of Melbourne, VIC 3010, Australia}
\newcommand{\sydney}{School of Physics, The University of Sydney, NSW 2006 Camperdown, Sydney, Australia}
\newcommand{\bern}{Albert Einstein Center for Fundamental Physics, Institute for Theoretical Physics, University of Bern, Sidlerstrasse 5, 3012 Bern, Switzerland}
\newcommand{\barcelona}{Department of Quantum Physics and Astrophysics and Institute of Cosmos Sciences, University of Barcelona, 08028 Barcelona, Spain}
\newcommand{\alabama}{Department of Physics and Astronomy, University of Alabama, Tuscaloosa, AL 35487, USA}
\newcommand{\darmstadt}{Department of Physics, Technische Universita\"at Darmstadt, 64289 Darmstadt, Germany}
\newcommand{\sissa}{Theoretical and Scientific Data Science, Scuola Internazionale Superiore di Studi Avanzati (SISSA), 34136 Trieste, Italy}
\newcommand{\dresden}{Technische Universit\"at Dresden, 01069 Dresden, Germany}
\newcommand{\alsoatcoimbrapoli}{Coimbra Polytechnic - ISEC, 3030-199 Coimbra, Portugal}
\newcommand{\alsoatroma}{INFN - Roma Tre, 00146 Roma, Italy}
\newcommand{\alsoatgrenada}{University of Grenada}
\newcommand{\alsoatbanjaluka}{University of Banja Luka, 78000 Banja Luka, Bosnia and Herzegovina}
\newcommand{\alsoatmpik}{Max-Planck-Institut f\"ur Kernphysik, 69117 Heidelberg, Germany}
\authorrunning{DARWIN Collaboration}
\thankstext{addr46}{Also at \alsoatbanjaluka}
\thankstext{addr44}{Also at \alsoatroma}
\thankstext{addr43}{Also at \alsoatcoimbrapoli}
\thankstext{addr45}{Also at \alsoatgrenada}
\thankstext{addr47}{Also at \alsoatmpik}

\thankstext{email2}{\texttt{jose.cuenca@physik.uzh.ch}}
\thankstext{email1}{\texttt{darwin-pub@darwin-observatory.org}}
\institute{\zurich\label{addr17}
\and
\munster\label{addr7}
\and
\lpnhe\label{addr18}
\and
\torino\label{addr14}
\and
\nikhef\label{addr8}
\and
\belgrade\label{addr33}
\and
\columbia\label{addr3}
\and
\nyuad\label{addr9}
\and
\alabama\label{addr38}
\and
\kitipe\label{addr28}
\and
\melbourne\label{addr34}
\and
\subatech\label{addr13}
\and
\bologna\label{addr0}
\and
\mpik\label{addr6}
\and
\kitiap\label{addr26}
\and
\sydney\label{addr35}
\and
\wis\label{addr16}
\and
\freiburg\label{addr19}
\and
\coimbra\label{addr2}
\and
\heidelberg\label{addr31}
\and
\stockholm\label{addr12}
\and
\lngs\label{addr4}
\and
\mainz\label{addr5}
\and
\kitetp\label{addr27}
\and
\rice\label{addr11}
\and
\laquila\label{addr22}
\and
\purdue\label{addr10}
\and
\chicago\label{addr1}
\and
\nagoya\label{addr21}
\and
\bern\label{addr36}
\and
\napels\label{addr20}
\and
\ucsd\label{addr15}
\and
\heidelbergki\label{addr32}
\and
\tokyo\label{addr23}
\and
\barcelona\label{addr37}
\and
\kobe\label{addr24}
\and
\sissa\label{addr40}
\and
\darmstadt\label{addr39}
\and
\ferrara\label{addr30}
\and
\dresden\label{addr41}
}

\maketitle

\begin{abstract}
Xenon dual-phase time projections chambers (TPCs) have proven to be a successful technology in studying physical phenomena that require low-background conditions. With $\SI{40}{t}$ of liquid xenon (LXe) in the TPC baseline design, DARWIN will have a high sensitivity for the detection of particle dark matter, neutrinoless double beta decay ($0\upnu\upbeta\upbeta$), and axion-like particles (ALPs). Although cosmic muons are a source of background that cannot be entirely eliminated, they may be greatly diminished by placing the detector deep underground. In this study, we used Monte Carlo simulations to model the cosmogenic background expected for the DARWIN observatory at four underground laboratories: Laboratori Nazionali del Gran Sasso (LNGS), Sanford Underground Research Facility (SURF),  Laboratoire Souterrain de Modane (LSM) and SNOLAB. We determine the production rates of unstable xenon isotopes and tritium due to muon-included neutron fluxes and muon-induced spallation. These are expected to represent the dominant contributions to cosmogenic backgrounds and thus the most relevant for site selection.
\end{abstract}

\section{Introduction}
\label{sec:intro}

Dual-phase xenon TPCs hold the best constraints for direct detection of weakly interacting massive particles (WIMPs) for masses above $\SI{6}{GeV/c^2}$\cite{XENON:DMresults,LUX:DMresults}, and they have increased the current sensitivity with their upgraded versions \cite{XENON:WIMPsens,LUX-ZEPLIN:WIMPsens, PandaX:comm}. Likewise, experiments using xenon in its pure form or dissolved in a liquid scintillator provide competitive limits on the half-life of the neutrinoless double beta decay of \ce{^{136}Xe} \cite{EXO-200:2019rkq, KamLAND-Zen:2016pfg, NEXT:2015rel}, with plans to probe half-lives up to two orders of magnitude larger in future and upgraded versions of these experiments \cite{nEXO:2017nam,Dolinski:2019nrj,Gomez-Cadenas:2019sfa}. The latest achievements in background mitigation, together with the increase of the target masses, have demonstrated that xenon experiments, and in particular xenon TPCs, are powerful instruments for other rare-event searches as well. This includes the observation of the extremely rare decay of \ce{^{124}Xe} via double electron capture \cite{XENON:2019dti}, the search for solar neutrino interactions on nuclei \cite{XENON:2020gfr}, and searches for solar axions, axion-like particles and dark photons \cite{LUX:2017glr, XENON:2020rca}. 

DARWIN, a proposed next-generation xenon experiment, will push the sensitivity to all these phenomena even further \cite{DARWIN:2016hyl,DARWIN:2020bnc,DARWIN:2020jme}. In its baseline design, DARWIN will use $\SI{40}{t}$ of instrumented liquid xenon in a dual-phase xenon TPC to complete an extensive science program. To fully exploit the physics goals of DARWIN an unprecedented, ultra-low background level will be required. To achieve such a level, external radiogenic backgrounds from detector materials as well as the concentration of \ce{^{222}Rn}, which emanates from detector components, have to be reduced to negligible contributions. In the ideal scenario, the detector is dominated by irreducible backgrounds such as neutrino interactions or decays of xenon isotopes present in the natural xenon. With this, cosmogenic activation of detector materials can turn out to be a dominant source of backgrounds. 

Even though the detectors are placed deep underground to shield them from cosmic radiation, muons with an energy $\mathcal{O}{(}\SI{100}{GeV}{)}$ can penetrate several kilometers of rock, concrete and shielding materials and reach the detector, producing a considerable amount of secondary neutrons through hadronic and electromagnetic showers that can potentially mimic a WIMP nuclear recoil interaction. These muon-induced neutrons can also produce long-lived radionuclides via different processes, such as inelastic interactions or captures. Radionuclides produced in the xenon volume as a consequence of these processes are a constant and irreducible source of background events in a wide energy range that can turn dominant if all the other background contributions are negligible. Additional veto systems, such as muon-veto water-filled tanks (Sec. \ref{subsec:muVeto}), help suppress the flow of incident neutrons onto the detector. Moreover, activation of the xenon above ground during production, storage and transportation, can be problematic too \cite{Baudis:2015kqa}. However, this can be mitigated by taking proper actions before construction and filling of the detector.

One of the most significant contributions to the background in neutrinoless double beta decay experiments using xenon detectors is the decay of \ce{^{137}Xe} which is produced after a neutron is captured by \ce{^{136}Xe}. This isotope $\upbeta$-decays to \ce{^{137}Cs} with a $Q_{\beta}$ of $\SI{4.162}{MeV}$ while the neutrinoless double beta decay of \ce{^{136}Xe} has a $Q_{\beta\beta}$ of $\SI{2.458}{MeV}$. Since the electrons produced in the \ce{^{137}Xe} decay have a continuous energy spectrum, they can potentially mimic the signal of a neutrinoless double beta decay, therefore reducing the sensitivity of the detector to this process. Experiments devoted to the neutrinoless double beta decay searches using \ce{^{136}Xe}, such as nEXO, KamLAND-Zen or NEXT have estimated a significant \ce{^{137}Xe} background \cite{NextCosm,nEXO:2017nam,Kamland23}. 

For the cosmogenic activation during operation, the depth of the underground laboratory becomes a crucial factor: the deeper the experimental site, the lower the muon flux, but, at the same time, the higher the mean energy of the muons and of their induced secondary particles. Given that the underground location of the DARWIN experiment is still to be decided, the study of the production of cosmogenic backgrounds at different sites will inform this choice. 

An accurate evaluation of the cosmogenic background requires realistic simulations. Using several already existing simulation packages, we developed a framework to perform full Monte Carlo simulations with muons at several underground locations.

This work is organized as follows: In Sect.~\ref{sec:structure} we describe the simulation framework. Sect.~\ref{sec:cosmoneutrons} focuses on the most problematic and abundant particles produced due to muon interactions, the muon-induced neutrons. In Sect.~\ref{sec:xe137} and Sect.~\ref{sec:tritium} we discuss the production of two radionuclides that can affect the sensitivity of DARWIN to different physics channels. Finally, we discuss and summarize the main results of our simulations in Sect.~\ref{sec:conclusions}.

\section{The DARWIN-Geant4 simulation framework}
\label{sec:structure}

The transport code \geant{} \cite{Geant42003} is one of the most robust simulation tools currently available and its use is very common in high energy physics. We have developed the DARWIN-Geant4 software package to perform the entire set of simulations for our detector. This framework is constructed using the libraries distributed with the \texttt{10.6.p02} version of \geant{}.

In the next subsections, we enumerate the geometry components that make up the DARWIN experiment (Subsec. \ref{subsec:halls}, \ref{subsec:muVeto}, \ref{subsec:tpc}). This includes the shielding materials of the experimental hall (rock and concrete) together with the veto systems and the TPC. Then, we list the set of physics processes that were used in these simulation  (Subsec. \ref{subsec:physList}). Finally, we describe the working principle of the muon generator tool and how it is adapted to the different underground locations (Subsec. \ref{subsec:muGen})

\subsection{Simulation of the experimental hall}
\label{subsec:halls}

\begin{figure}[h!]
	\centering
	\includegraphics[width=\linewidth]{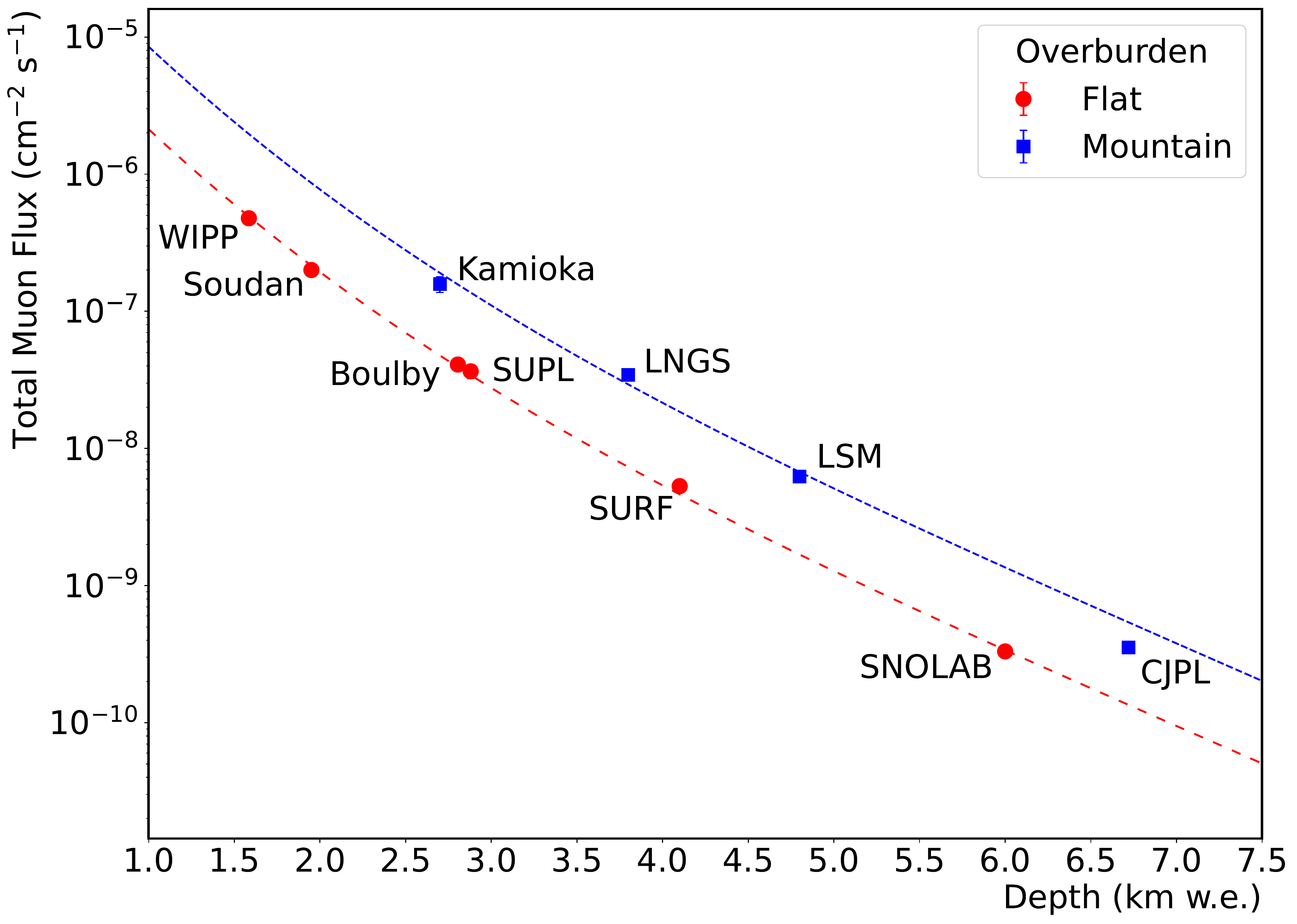}
	\caption{Total muon flux for several underground laboratories as a function of the depth in kilometers water equivalent (km w.e.). The lines are obtained using empirical fits as explained in \cite{Mei2006, Guo2021}. The numerical values for the laboratories studied in this work are shown in Table \ref{tab:muonFlux}.}\label{fig:muonLabs}
\end{figure}

Each site considered in this study has advantages and disadvantages that must be addressed, regarding not only physics but also logistics, etc. For this work, the parameters that we considered relevant were the size of the experimental hall, the measured total muon flux, the mean energy (that is related to the depth of the site) and the angular distribution of the incident muons. In Fig. \ref{fig:muonLabs} the measured total muon flux as a function of the depth for several underground laboratories is shown.

Some of these laboratories have several experimental halls that could be potentially available. However, since other experiments are underway, we have selected those halls that can be best adapted to the size of the DARWIN baseline design and according to our scheduled timeline. Based on the size of the hall, the number and size of components in the experimental setup may be reduced or modified, especially in the case of the muon veto system (Sec. \ref{subsec:muVeto}). The four laboratories that were considered in this study are LNGS, SURF, LSM and SNOLAB. In order to perform a realistic simulation, we have implemented the geometry of the experimental halls in our simulation framework using the technical drawings provided by the laboratories.

Since all these facilities are shielded by several kilometers of rock, it is necessary to determine how much surrounding material needs to be simulated. To ensure an equilibrium between the secondary particles production and muon fluxes, we used $\SI{5}{m}$ of rock for the walls of the halls. This value has been shown in previous cosmogenic simulations to be sufficient to achieve this equilibrium \cite{Selvi2011}, and we used it in all four locations of this study. Besides the rock, other shielding materials, such as concrete, may be also present. The additional layers of concrete were simulated following the specifications indicated by the technical drawings. 

We now briefly describe the four laboratories, including the values of the geometrical parameters that were implemented in our simulation framework.

\subsubsection*{LNGS}
\label{subsub:GS}

Laboratori Nazionali del Gran Sasso (LNGS) are located in Italy and hosts, among others, the XENONnT dark matter experiment. It has three main experimental halls and for these simulations, we have implemented Hall B. Its $xy$-projection is a rectangle of $\SI{18.2}{m}\; \times \; \SI{60.0}{m}$ and the ceiling is a cylindrical vault with a maximum height of $\SI{20}{m}$. The inner part of the hall is covered with a layer of concrete of $\SI{50}{cm}$ thickness. The chemical composition that we implemented in the code for the rock and concrete materials is defined in \cite{wulandari2004}.

\subsubsection*{SURF}
\label{subsub:SURF}

Sanford Underground Research Facility (SURF) is a former gold mine located in South Dakota and currently is hosting the LZ dark matter experiment. In our code, we simulated the Davis cavern as a box of $\SI{17}{m\,(l)}\; \times \; \SI{10}{m\,(w)}\; \times\; \SI{12}{m\,(h)}$. It has no concrete layer, and therefore the entire hall is made of rock. The chemical composition of the SURF rock is implemented as in \cite{MeiSurf2010}.

\subsubsection*{LSM}
\label{subsub:LSM}

In France, Laboratoire Souterrain de Modane (LSM) has also hosted dark matter experiments such as EDELWEISS. The only available place is the Grand Hall, with an area of $\SI{10}{m}\; \times \; \SI{20}{m}$. The ceiling is an elliptical dome that provides a maximum height of approximately $\SI{17}{m}$. The inner part of the hall is covered by a $\SI{30}{cm}$ layer of concrete. The chemical composition of rock and concrete was implemented as in \cite{Kluck2015}.

\subsubsection*{SNOLAB}
\label{subsub:SNOLAB}

The deepest location considered in this study is SNOLAB. It is located in Canada and in the future it will host the SuperCDMS dark matter experiment. For this work, we simulated the Cube hall. It is a box $\SI{18}{m\,(l)}\; \times\; \SI{15}{m\, (w)}\; \times\; \SI{20}{m\, (h)}$. For the shielding materials we used the composition in the SNOLAB reference manual \cite{SnolabManual}.

\subsection{The muon veto system}
\label{subsec:muVeto}

In the DARWIN baseline design, the LXe TPC is located inside of a water tank that serves as water-Cherenkov muon veto. This water tank is implemented here as a $\SI{0.5}{cm}$ thick stainless steel cylinder with a truncated cone on top filled with water. For the cylinder of radius $R$ and height $h$, the top cone has a height of $h/5$ and a top radius of $R/4$. 

In the baseline design, the water tank has a radius of $\SI{6}{m}$. However, a tank of this size does not fit in all underground locations of this study. We adapted the size of the tank where necessary while preserving the aspect ratio, as indicated in Table \ref{tab:wt}. The muon veto influences the cosmogenic background rate both by moderating neutrons produced in the rock wall, and by actively vetoing muons with trajectories that pass the TPC.

\begin{table}[!htbp]
\centering
\caption{Summary of the geometrical parameters implemented for the underground laboratories of this study. The second column shows the names of the experimental hall, together with the shape of the ceiling (third column). Fourth and fifth columns show the radius and total height of the water tank simulated for each hall.}\label{tab:wt}
\begin{tabular}{lllcc}
\hline
Site  & Hall & Shape & R$_{wt}$ (m)  & h$_{wt}$ (m) \\
\hline
LNGS & B            & Vault &  6  & 12.0 \\
SURF  & Davis Cavern & Box   &  4  & 9.6  \\
LSM   & Grand Hall   & Vault &  4  & 9.6  \\
SNOLAB    & Cube Hall    & Box   &  6  & 14.4 \\
\hline
\end{tabular}
\end{table}

\subsection{The DARWIN cryostat and TPC}\label{subsec:tpc}

\begin{figure}[h!]
	\centering
	\includegraphics[width=\linewidth]{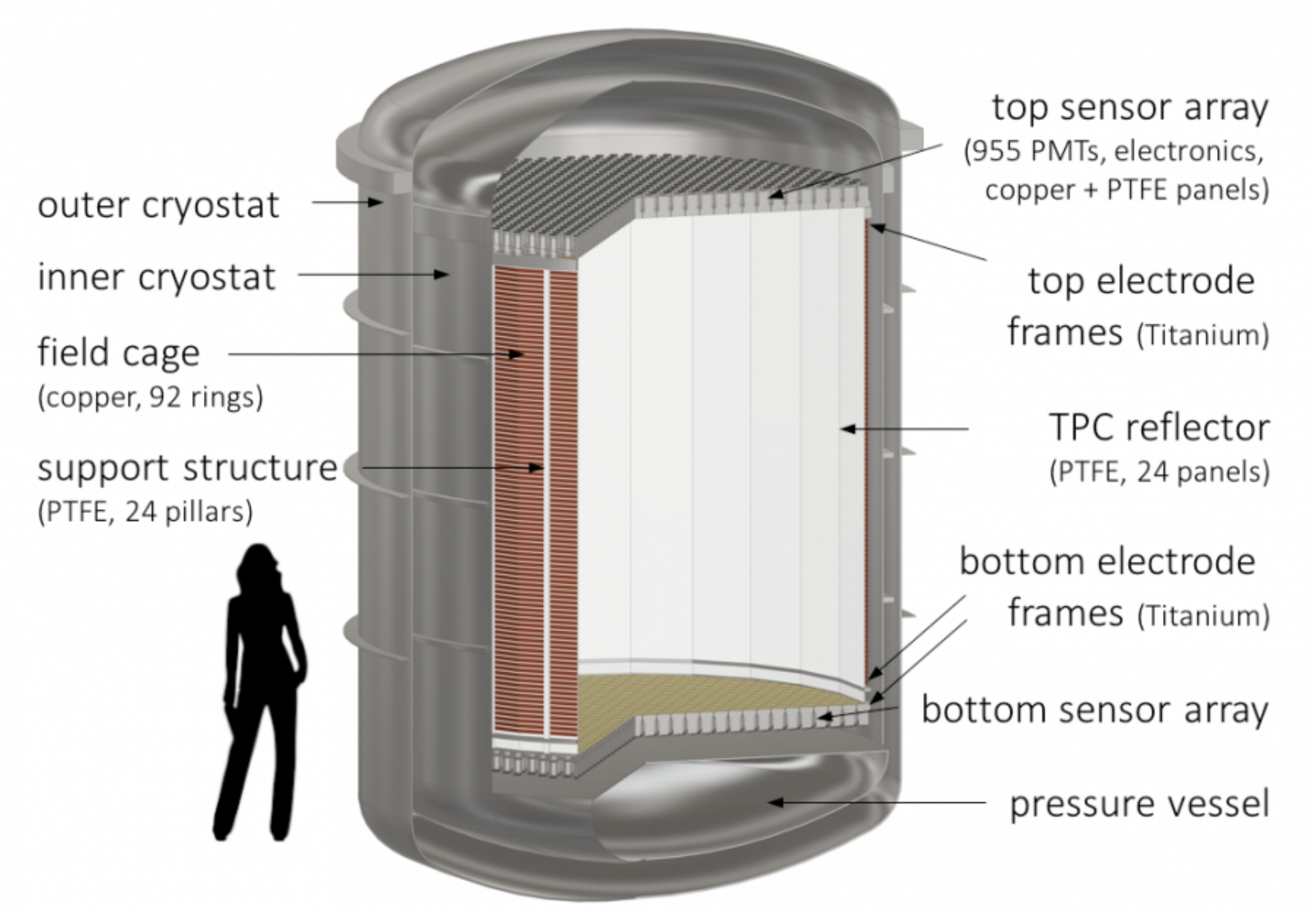}
	\caption{Schematic overview of the DARWIN geometry implemented in the DARWIN-Geant4 framework. It consists of a cylindrical TPC, with two sensor arrays, electrodes and field shaping rings surrounding the active volume. The TPC is placed inside a double walled cryostat made of titanium.}\label{fig:tpc}
\end{figure}

For the simulations, we have used the same detector geometry as implemented in \cite{DARWIN:2020jme}. In the baseline design, the DARWIN TPC consists of a cylindrical dual-phase TPC with a diameter and height of $\SI{2.6}{m}$ (aspect ratio 1:1), see Fig. \ref{fig:tpc}. It has two photosensor arrays (on top and bottom) formed by 995 \texttt{R11410-21} Hamamatsu photomultiplier tubes (PMTs)\cite{Baudis2013}. The PMTs are held in place by polytetrafluoroethylene (PTFE) and copper disks. The lateral part of the cylinder is fully covered with PTFE reflector panels and 24 supporting PTFE pillars. The detector is equipped with a set of electrodes and 92 copper field cage surrounding the TPC.

The TPC is placed inside a double-walled vacuum insulated cryostat with a wall thickness of $\SI{5}{mm}$ made of titanium. The vessels are simulated as cylindrical bodies finished with torus-spherical domes (DIN 28011 \cite{DIN28011}) on top and bottom. Stiffener rings are also added to the design in order to prevent deformations of the vessel due to the pressure. Additionally, a filler vessel, consisting of a metallic shell filled with pressurized gas, is positioned in the bottom dome to yield a flat bottom surface and minimize the liquid xenon filling mass. 

The active volume is filled with a total mass of $\SI{40}{t}$ of natural liquid xenon together with $\SI{30}{kg}$ of gaseous xenon on the top part of the TPC and the upper dome. For the simulation of the active xenon material, we have used the isotopic composition of natural xenon provided by the \geant{} material database. The mixture, together with the abundance of each isotope, is detailed in Table \ref{tab:xeIso}.

\begin{table}[!htbp]
\centering
\caption{Natural abundance of xenon isotopes used in these simulations. The values are given by the \geant{} internal material database (NIST).} 
\label{tab:xeIso}
\begin{tabular}{lcc}
\hline
Isotope  & Abundance (\%)  \\
\hline
\ce{^{124}Xe} & 0.09 \\
\ce{^{126}Xe} & 0.09 \\
\ce{^{128}Xe} & 1.92  \\
\ce{^{129}Xe} & 26.44  \\
\ce{^{130}Xe} & 4.08 \\
\ce{^{131}Xe} & 21.18  \\
\ce{^{132}Xe} & 26.89  \\
\ce{^{134}Xe} & 10.44  \\
\ce{^{136}Xe} & 8.87  \\
\hline
\end{tabular}
\end{table}

\subsection{Geant4 physics lists}
\label{subsec:physList}

The \geant{} toolkit distributes a set of physics lists that are maintained and updated with each version. A full detailed description of these physics lists can be found in the \geant{} physics reference manual \cite{GeantPhys}, together with their recommended usage. 

We incorporated the modular nature of \geant{} in our simulation framework, which allows the user to create custom physics lists. We have carried out a systematic study of the features and consistency of several of them by means of toy Monte Carlo simulations. These simulations had a very simplified geometry formed by a rectangular block of rock of $\SI{5}{m}$ thickness and a cylinder of $\SI{2}{m}$ radius, $\SI{2}{m}$ height filled with liquid xenon. The primary particles were vertical muons with fixed energy of $\SI{300}{GeV}$. As control parameters, we have chosen the muon-induced neutrons produced in the rock and the \ce{^{137}Xe} production rate.

The physics lists that we used in these toy simulations were implemented in two different ways: \emph{pure} and \emph{mixed} configurations. In the pure version, we used the physics lists as they are distributed with our \geant{} release. In this configuration, we tested the \texttt{Shielding}, \texttt{ShieldingLEND}, \texttt{QGSP\!\_BIC\!\_HP} and \texttt{QGSP\!\_BERT\!\_HP} lists. For the mixed configurations, we use the \texttt{emlivermore} physics list for the electromagnetic processes, while for the hadronic interactions we used the hadronic part of the \texttt{Shielding}, \texttt{ShieldingLEND}, \texttt{QGSP\!\_BIC\!\_HP} and \texttt{QGSP\!\_BERT\!\_HP} physics lists. These physics lists have been validated in simulations of experiments placed underground, being the \texttt{Shielding} list the recommended by the Geant4 developers in these type of simulations. The \texttt{ShieldingLEND} list is a flavour of \texttt{Shielding} that uses the Low-Energy Nuclear Database (LEND) database to describe the transport of low-energy neutrons.

The results produced by the toy simulations did not show a significant discrepancy in the values of the control parameters. All results were of the same order of magnitude and they were compatible within a factor $0.7$ in the \ce{^{137}Xe} production rate. We attribute this to the difference between the models implemented in the physics lists. Even though the pure configurations produced similar results as the mixed, the computational time of the second is significantly larger than the first. In this set of toy simulations, we estimated that simulations using the pure configurations are $\sim 20\%$ faster.

Therefore, we decided to perform the full simulations using the pure \texttt{Shielding} as the main list together with the pure \texttt{ShieldingLEND} and \texttt{QGSP\!\_BIC\!\_HP} lists as control simulations. 

\subsection{The primary muon generator}
\label{subsec:muGen}

We used the MUSIC-MUSUN software \cite{Kudryavtsev2009} to obtain the kinematic information of muons that serves as input to a custom muon generator in \geant{}. With this software it is possible to simulate the energy-direction correlation of the muons at each underground facility taking into account parameters such as rock density or the orography of the lab. The input consists of a mixture of $\upmu^{+}$ and $\upmu^{-}$ muons with a ratio of $\upmu^{+}/\upmu^{-}=1.3$ in their populations. A summary of the properties of the primary muons is found in Table \ref{tab:sims}.

\begin{table}[!htbp]
\centering
\caption{Mean values of the energy and angles (direction) of the primary muons produced with the MUSUN software for the underground locations considered in this study. The right column shows the density of the shielding rock used in MUSUN and in our \geant{} code.}
\label{tab:sims}
\begin{tabular}{lcccc}
\hline
      & $\langle E_{\mu} \rangle$ & $\langle \theta \rangle$ & $\langle \Phi \rangle$ & $\langle \rho \rangle$ \\
Site  & ($\SI{}{GeV}$) & ($\SI{}{deg}$) & ($\SI{}{deg}$) & ($\SI{}{g\per cm^{3}}$) \\
\hline
LNGS & 272.7 & 37.42 & 200.90  &  2.71 \\
SURF  & 284.7 & 27.09 & 170.38  &  2.70 \\
LSM   & 301.3 & 37.55 & 169.90  &  2.65 \\
SNOLAB    & 309.6 & 24.69 & 179.96  &  2.83 \\
\hline
\end{tabular}
\end{table}

For each event, the generator reads a line of the MUSUN file containing the muon-type ($\upmu^{+}$ or $\upmu^{-}$), its energy and the direction of propagation. The position of the primary vertex is found using a three-step algorithm that we call \emph{random sampling-rotation-projection}. In this algorithm we use a disk of radius $R$ centred at the origin of the coordinate system, which is set at the center of the TPC. This algorithm, is analogous to the sampling methods used in \cite{Selvi2011,Kluck2015} and works as follows: 
\begin{enumerate}
    \item Random sampling. A random point is generated inside a disk of radius $R$ on the horizontal $z=0$ plane. We transport the direction vector to this point.  
    \item Rotation. We rotate the disk on $z=0$ until it becomes perpendicular to the direction vector.
    \item Projection. The point on the rotated disk is projected to the external surface of the rock volume, assuming a linear trajectory.
\end{enumerate}

The radius $R$ of the sampling disk has to be large enough to cover the whole experimental setup, but at the same time not excessively large, as a very large disk could imply that we are sampling muons that have no physical meaning or that are so far away from the experiment that they are irrelevant for the simulation. Since the live-time of the simulation is estimated from the size of the disk, a large disk translates into more computational time needed to produce significant statistics. Since every muon has its initial position at the external surface of the shielding materials, the muon generator ensures that all the primary muons are propagated through at least $\SI{5}{m}$ of rock. 

In Table \ref{tab:muonFlux} we summarised the experimentally measured muon fluxes at the locations of this work together with the radius of the sampling disk used in the muon generator and the equivalent live-time of the simulations.

\begin{table}[!htbp]
\centering
\caption{Measured muon fluxes, $\phi\pm\sigma(\phi)$, for the underground laboratories considered in this study. The third column shows the radius of the disk used in the muon generator and the last column is the equivalent live-time.}
\label{tab:muonFlux}
\begin{tabular}{lcccl}
\hline
Site       & Muon flux ($\SI{}{cm^{2}\,s^{-1}}$)  & $R$ ($\SI{}{m}$) & $T$ ($\SI{}{yr}$)\\
\hline
LNGS & $(3.432 \pm 0.003)\cdot 10^{-8}$\cite{GsMuon2019}       &  10  & 29.41 \\
SURF  & $(5.31  \pm 0.17) \cdot 10^{-9}$\cite{SurfMuon2017}     &   8  & 29.70 \\
LSM   & $(6.25  \pm 0.23) \cdot 10^{-9}$\cite{LsmMuon2013}      &   6  & 44.86 \\
SNOLAB    & $(3.31  \pm 0.09) \cdot 10^{-10}$\cite{SnolabMuon2009}  &   9  & 232.93 \\
\hline
\end{tabular}
\end{table}

\section{Results}

\subsection{Muon-induced neutrons}
\label{sec:cosmoneutrons}

Muons produce cascades of secondary particles when they interact with the different materials of the experiment. A full description of such cascades is complicated due to the types of particles and the diversity of physical processes involved. The interaction of cosmic muons with the shielding rock of the experiment is the main source of environmental neutrons. The rate of these neutrons, for the underground locations considered in this study, is about one order of magnitude smaller compared to the neutrons produced in the spontaneous fission and $(\upalpha,n)$ reactions of \ce{^{238}U} and \ce{^{232}Th} present in the rock and concrete. However, muon-induced neutrons can reach energies up to several $\SI{}{GeV}$. It is thus very difficult to completely stop them and, therefore, they can be a potential background source inside the detector.  

\begin{figure}[h!]
	\centering
	\includegraphics[width=\linewidth]{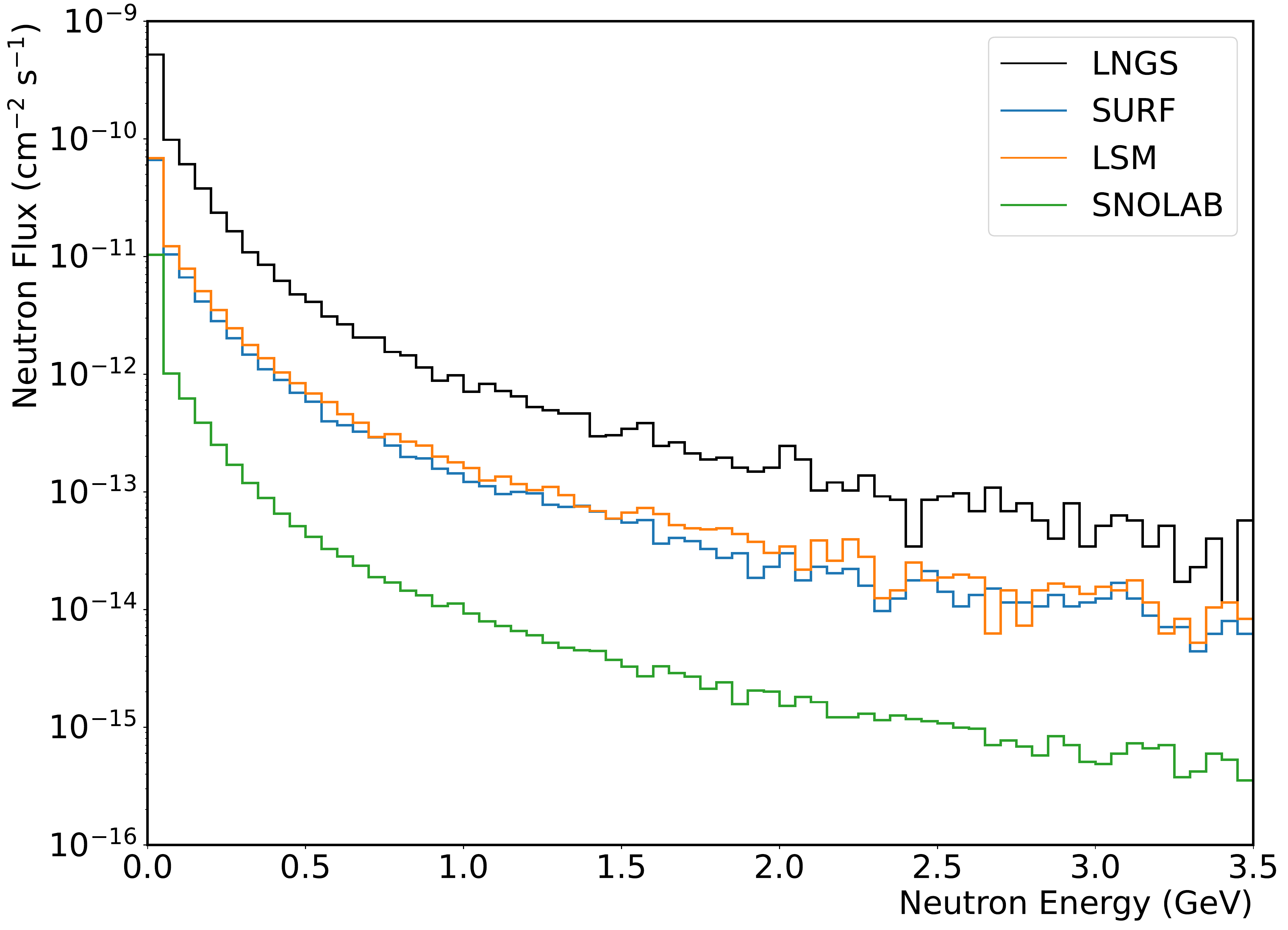}
	\caption{Spectrum of the muon-induced neutrons produced in rock (and concrete) for the underground laboratories considered in this work. The size of the energy bins is $\SI{50}{MeV}$.}\label{fig:muonNeutrons}
\end{figure}

Fig. \ref{fig:muonNeutrons} shows the energy spectrum of the muon-induced neutrons obtained in our simulations for the four underground laboratories using the \texttt{Shielding} physics list. The neutrons shown in the figure are the neutrons that enter the experimental hall from the rock and concrete walls. The double counting of the neutrons, such as neutrons bouncing off the walls, was avoided by a careful selection based on the track information provided by \geant{}. For these, only the first step with the initial energy of the neutron entering the laboratory is considered. 

Past versions of \geant{} underestimated the production of these neutrons. The simulations done in \cite{Selvi2011} studying the efficiency of the muon veto system for the XENON1T experiment show a muon-induced neutron rate an order of magnitude smaller compared to similar simulations done with the FLUKA simulation package \cite{Mei2006}. This large discrepancy suggested that the neutron yields calculated with \geant{} (version 9.3) were systematically underestimated due to the lack of implementation of the cross sections of the muon interactions. This issue is not noticed in the \geant{} version \texttt{10.6.p2}.

Table \ref{tab:integralFlux} shows the integrated values of the neutron fluxes for different energy thresholds estimated in this work compared to the FLUKA simulations in \cite{Mei2006}. 

\begin{table}[!htbp]
\centering
\caption{Integrated muon-induced neutron flux (in $10^{-9} \SI{}{cm^{-2}\, s^{-1}}$) for the underground laboratories considered in this study using the \texttt{Shielding} physics list. The values in brackets are the FLUKA simulations in \cite{Mei2006}.}
\label{tab:integralFlux}
\begin{tabular}{llll}
\hline
Site       & E $> \SI{1}{MeV}$  &  E $> \SI{10}{MeV}$ & E $> \SI{100}{MeV}$ \\
\hline
LNGS & 0.83 (0.81)   &  0.37 (0.73)    & 0.14 (0.201)   \\
SURF  & 0.15          &  0.067          & 0.024          \\
LSM   & 0.16          &  0.071          & 0.027          \\
SNOLAB    & 0.015 (0.020) &  0.005 (0.018)  & 0.0017 (0.005) \\
\hline
\end{tabular}
\end{table}

The integrated spectra differ around $2-3\,\%$  and they are of the same order of magnitude in the three energy ranges of interest. From these results, we conclude that the current version of \geant{} and the implementations done in the physics lists provide a similar description of the production of neutrons in rock and concrete as FLUKA. 

\subsection{Production of \ce{^{137}Xe}}
\label{sec:xe137}

For all the rates presented in this work, the value obtained by the \texttt{Shielding} physics list is the final production rate, and the systematic uncertainty is estimated using the other two:
\begin{equation}
    \sigma_{sys}(R_{sh})=\sqrt{\frac{(R_{sh}-R_{shL})^{2} + (R_{sh}-R_{bic})^{2}}{2}}
\end{equation}
where $R_{sh}$, $R_{shL}$, and $R_{bic}$ are the rates obtained with \texttt{Shielding}, \texttt{ShieldingLEND}, and \texttt{QGSP\!\_BIC\!\_HP} respectively.

For the total uncertainty in the rate, two extra components are considered: first, the uncertainty in the number of isotopes (statistical) and second the experimental uncertainty of the measured muon flux. The first is assumed to be Poisson-like distributed and therefore the associated uncertainty is $\sqrt{n}$, being $n$ the number of produced isotopes. The uncertainty in the muon flux translates into an uncertainty in the simulated live-time ($T$). Both components are then added in quadrature to provide the total uncertainty of the rate:
\begin{equation}
    \sigma(R_{sh}) = R_{sh}\cdot \sqrt{\frac{1}{n}+\frac{\sigma^{2}(\phi)}{\phi^{2}}} \; ,
\end{equation}
where $n$ is the number of isotopes and $\phi$ the measured muon flux with uncertainty $\sigma(\phi)$ as seen in Table \ref{tab:muonFlux}.

\begin{figure}[h!]
	\centering
	\includegraphics[width=\linewidth]{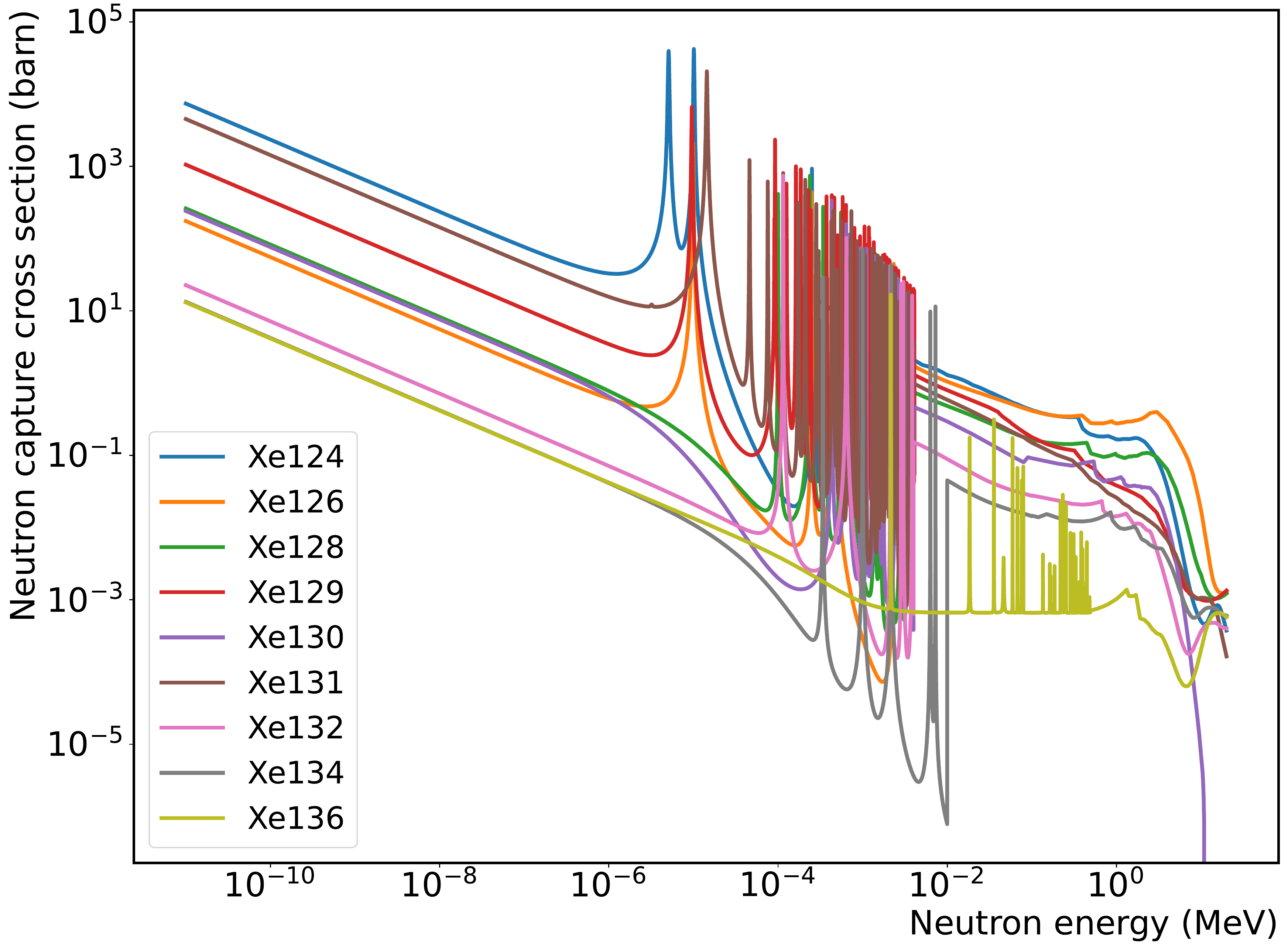}
	\caption{Neutron capture cross sections for the xenon isotopes as a function of the neutron energy, taken from \cite{exfor}.}\label{fig:XeCrossSec}
\end{figure}

Fig. \ref{fig:XeCrossSec} shows the cross section for the neutron capture process in xenon isotopes as a function of the incident neutron energy.

Despite the neutron capture cross section being higher for low energy neutrons, there are resonances for energies below $\SI{1}{MeV}$ and therefore neutrons that are not fully thermal can be captured too. In addition, since \ce{^{136}Xe} has the smallest neutron capture cross section (olive curve), the amount of neutrons available to be captured by \ce{^{136}Xe} is reduced by the presence of the other xenon isotopes. Further studies are being performed to study the cosmogenic \ce{^{137}Xe} production rate as a function of the isotopic composition of xenon at several underground locations. This new study will give more information on whether using enriched-depleted xenon in DARWIN is more convenient for the scientific channels of interest.

Our simulations indicate that the muon-induced neutrons produced in the walls of the laboratory are properly thermalized and captured by the muon veto water tank and they do not contribute to the production of \ce{^{137}Xe}. We conclude that approximately $95\%$ of the \ce{^{137}Xe} isotopes are produced by the capture of neutrons produced by muons crossing the liquid xenon. The rest are produced by secondary neutron cascades originated in the detector materials. 

\begin{table}[!htbp]
\centering
\caption{Muon-induced \ce{^{137}Xe} production rate at the different underground laboratories. The central value is the rate obtained with the \texttt{Shielding} physics list and the systematic error is calculated using the complementary simulations with the \texttt{ShieldingLEND} and \texttt{QGSP\!\_BIC\!\_HP} physics lists.} 
\label{tab:xe137}
\begin{tabular}{ll}
\hline
Site  &  Rate ($\SI{}{kg^{-1} yr^{-1}}$) \\
\hline
LNGS    &  $(8.22\pm 0.27 \pm 1.00_{sys} ) \cdot 10^{-4}$  \\
SURF    &  $(1.42\pm 0.12 \pm 0.21_{sys} ) \cdot 10^{-4}$  \\
LSM     &  $(1.65\pm 0.11 \pm 0.30_{sys} ) \cdot 10^{-4}$  \\
SNOLAB  &  $(6.75\pm 0.60 \pm 1.00_{sys} ) \cdot 10^{-6}$  \\
\hline
\end{tabular}
\end{table}

Table \ref{tab:xe137} summarizes the production of \ce{^{137}Xe} for the considered underground locations. In \ref{app:tables} the complete set of rates obtained for all locations with the three physics lists are shown. From those tables we have a maximum discrepancy of a factor $\sim 20\%$ between rates, which is expected due to the different physical models used in the definitions of the lists.

The value obtained in this work for the \ce{^{137}Xe} production rate at LNGS is a factor $\sim 8$ smaller compared to previous DARWIN study of the sensitivity to the neutrinoless double beta decay \cite{DARWIN:2020jme}. With the result presented in this work, the contribution of the decay of \ce{^{137}Xe} to the neutrinoless double beta decay background is lower than the contribution of the solar \ce{^{8}B} neutrinos at all underground locations.

\subsection{Production of tritium}
\label{sec:tritium}

In 2020, the XENON collaboration reported an excess of events at energies below $\SI{7}{keV}$ \cite{XENON:2020rca}. One of the possible explanations is the $\upbeta$-decay of tritium atoms present in the active xenon inside the TPC. This decay has a $Q_{\beta}$ of $\SI{18.591}{keV}$ and a half-life of $\SI{12.3}{yr}$. It was reported that the excess would correspond to a tritium concentration of $(6.2\;\pm\;2.0)\cdot 10^{-25} \,\SI{}{mol/mol}$.    

Tritium is mainly produced by neutron inelastic scattering and muon spallation processes. Although emanation from materials was considered as the primary source of tritium, cosmic muons and their secondary induced neutrons could be a continuous source of tritium. Spallation reactions of the xenon isotopes and in the surrounding materials of the TPC could induce the presence of tritium in the sensitive volume. Table \ref{tab:tritium} summarizes the production of \ce{^{3}H} for the considered underground locations.

\begin{table}[!htbp]
\centering
\caption{Muon-induced \ce{^{3}H} production rate at the different underground laboratories. The central value is the rate obtained with the \texttt{Shielding} physics list and the systematic error is calculated using the complementary simulations with the \texttt{ShieldingLEND} and \texttt{QGSP\!\_BIC\!\_HP} physics lists.} 
\label{tab:tritium}
\begin{tabular}{ll}
\hline
Site  & Rate ($\SI{}{kg^{-1} yr^{-1}}$)\\
\hline
LNGS    &  $( 1.22 \pm 0.01 \pm 0.01_{sys} )\cdot 10^{-2}$ \\
SURF    &  $(1.98  \pm 0.01 \pm 0.04_{sys})\cdot 10^{-3}$ \\
LSM     &  $(2.44  \pm 0.01 \pm 0.04_{sys})\cdot 10^{-3}$ \\
SNOLAB  &  $(1.40  \pm 0.05 \pm 0.50_{sys})\cdot 10^{-4}$ \\
\hline
\end{tabular}
\end{table}

Assuming the worst case scenario, in which all the tritium that has been produced cannot be removed and it remains in the active volume, after one year at LNGS, we expect $\SI{11.6}{\ce{^{3}H}/ton}$ or the equivalent of $2.6\cdot 10^{-27}\,\SI{}{mol/mol}$. This value has a discrepancy of two orders of magnitude with the tritium concentration that could explain the XENON1T excess. In the analysis of low-energy electronic recoil data from the first science run of the XENONnT experiment, the excess observed in XENON1T disappeared \cite{xenon2022}.

\subsection{Activation of other xenon isotopes}
\label{sec:xeisot}

Neutron captures can produce xenon isotopes inside the TPC that are unstable. In the low-energy region (below $\SI{30}{keV}$), the decays of these isotopes could be relevant for the WIMP and other rare-event searches. For example, \ce{^{125}Xe} decays to \ce{^{125}I} with a half-life of $\SI{16.9}{hr}$. The gamma lines from the \ce{^{125}Xe} decay are above $\SI{200}{keV}$, hence they are not of interest for the WIMP analysis. However, the \ce{^{125}I} also decays to \ce{^{125}Te}, with a half-life of $\SI{59.4}{d}$. In this case, the three lines from the \ce{^{125}I} decay are below $\SI{100}{keV}$. They correspond to the atomic K-shell, L-shell, and M-shell with decreasing probability and produce peaks at $\SI{67.3}{keV}$, $\SI{40.4}{keV}$, and $\SI{36.5}{keV}$, respectively. In addition to neutron capture, other processes such as muon spallation or photodisintegration contribute to the formation of the isotopes mentioned above. The KamLAND collaboration recently published a list of isotopes produced after muon spallation that have a non-negligible impact on the ROI of the $0\upnu\upbeta\upbeta$ process \cite{kamlandSpallation}.

The isotopes considered in this study are \ce{^{125}Xe} (t$_{1/2}=\SI{16.9}{hr}$), \ce{^{127}Xe} (t$_{1/2}=\SI{36.4}{d}$), \ce{^{133}Xe} (t$_{1/2}=\SI{5.24}{d}$) and \ce{^{135}Xe} (t$_{1/2}=\SI{9.14}{hr}$). The activation rates of these isotopes for the underground laboratories are shown in Table \ref{tab:xeRate}. We observe an agreement between the three physics lists used in our simulations.

\begin{table*}[!htbp]
\centering
\caption{Muon-induced \ce{^{125}Xe}, \ce{^{127}Xe}, \ce{^{133}Xe} and \ce{^{135}Xe} production rates (all processes) in $\SI{}{kg^{-1} yr^{-1}}$ at the different underground laboratories. The central value is the rate obtained with the \texttt{Shielding} physics list and the systematic error is calculated using the complementary simulations with the \texttt{ShieldingLEND} and \texttt{QGSP\!\_BIC\!\_HP} physics lists.} 
\label{tab:xeRate}
\begin{tabular}{ccccc}
\hline
Isotope & LNGS & SURF & LSM & SNOLAB \\
\hline
\ce{^{125}Xe} &  $(2.28 \pm 0.02 \pm 0.50_{sys})\cdot 10^{-2}$ 
& $(3.09\pm 0.01 \pm 0.50_{sys}) \cdot 10^{-3}$ 
& $(4.33 \pm 0.17 \pm 0.80_{sys}) \cdot 10^{-3}$ 
& $(2.18 \pm 0.09 \pm 0.50_{sys})\cdot 10^{-4}$ \\
\ce{^{127}Xe} &  $(6.39 \pm 0.02 \pm 0.60_{sys})\cdot 10^{-2}$ 
& $(1.02\pm 0.03 \pm 0.80_{sys}) \cdot 10^{-2}$ 
& $(1.25\pm 0.05 \pm 0.12_{sys})\cdot 10^{-2}$ 
& $(6.38 \pm 0.21 \pm 0.70_{sys})\cdot 10^{-4}$ \\ 
\ce{^{133}Xe} &  $(1.16\pm 0.01 \pm 0.10_{sys})\cdot 10^{-1}$ 
& $(1.80\pm 0.05 \pm 0.02_{sys}) \cdot 10^{-2}$ 
& $(2.23 \pm 0.08 \pm 0.16_{sys}) \cdot 10^{-2}$ 
& $(1.20 \pm 0.04 \pm 0.07_{sys})\cdot 10^{-3}$ \\ 
\ce{^{135}Xe} &  $(4.63\pm 0.02 \pm 0.16_{sys})\cdot 10^{-2}$ 
& $(7.42\pm 0.25 \pm 0.28_{sys})\cdot 10^{-3}$ 
& $(9.10\pm 0.30 \pm 0.50_{sys})\cdot 10^{-3}$ 
& $(5.01 \pm 0.02 \pm 0.02_{sys})\cdot 10^{-4}$ \\
\hline
\end{tabular}
\end{table*}

\subsection{Activation of isotopes in the xenon storage system}
\label{sec:column}

Muon-induced neutrons coming from the walls of the laboratory are moderated in the water tank and they have no influence on the activation of isotopes such as \ce{^{137}Xe}. However, the xenon storage and purification systems are placed outside the protection provided by the water tank, and exposed to high-energy neutrons and susceptible to activation. Since the liquid xenon is circulating between the purification systems and the TPC, radioactive isotopes activated outside the detector can end inside the active volume. 

\begin{figure}[h!]
	\centering
	\includegraphics[width=\linewidth]{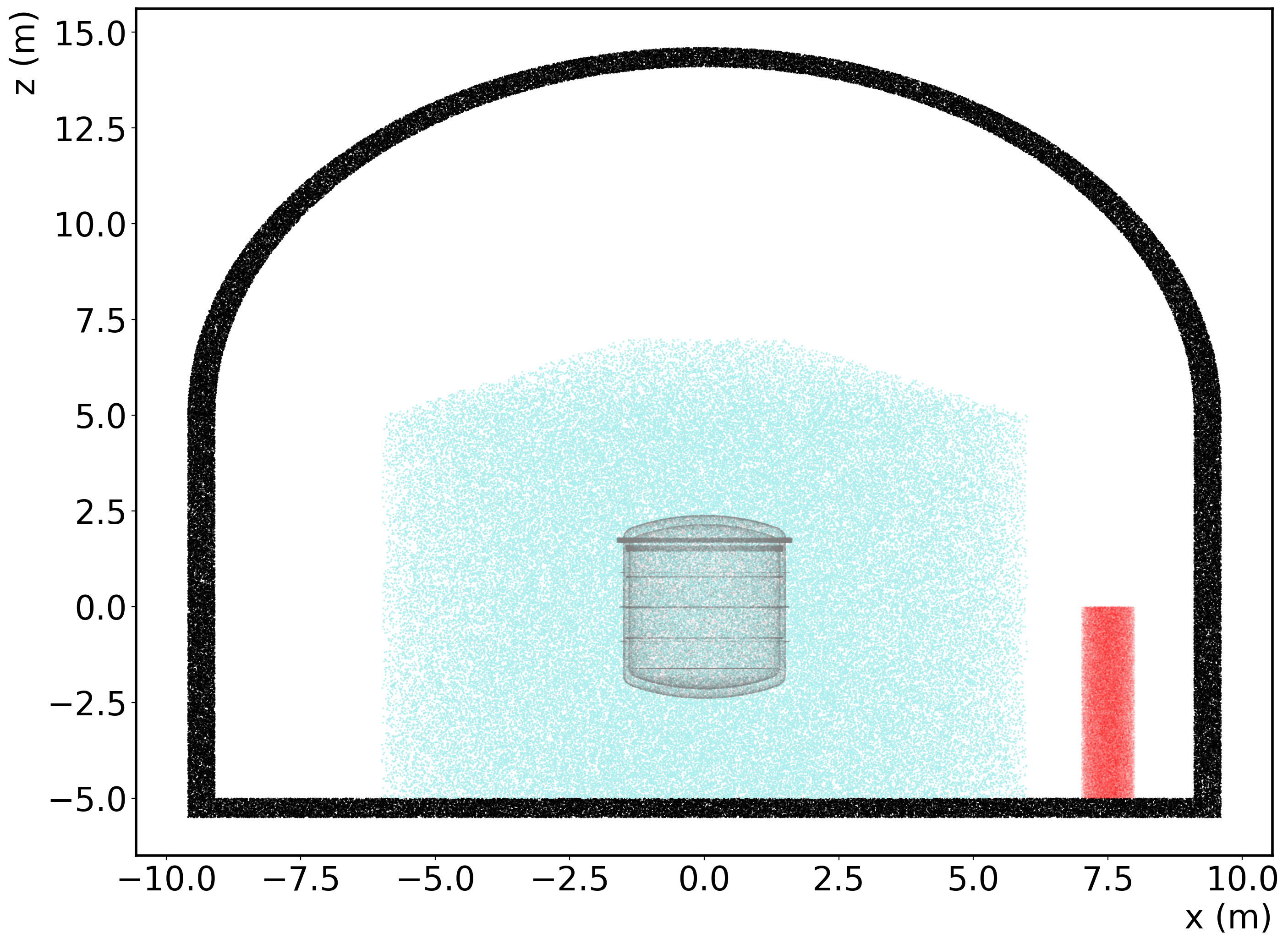}
	\caption{Cross-sectional view of the simulated LNGS experimental hall B with a storage column outside the water tank. The black region represents the concrete layer, the turquoise is the water tank, the grey is the cryostat and the red is the cylindrical storage column. The TPC and the shielding rock are removed for simplicity.\label{fig:column}}
\end{figure}

As the design of the DARWIN purification system is not finished yet, in a first approximation we simulated the storage system as a stainless-steel cylinder with a thickness of $\SI{0.5}{cm}$, $\SI{0.5}{m}$ radius, and $\SI{5}{m}$ height at the LNGS Hall B, as seen in Fig. \ref{fig:column}. In total, this column contains $\SI{11.2}{tonnes}$ of liquid xenon in addition to those already present inside the detector.

We simulated a total live-time of approximately $\SI{9.3}{yr}$ using the information for muons provided by MUSUN in our muon generator and the \texttt{Shielding} physics list. From this simulation, we obtain production rates for tritium and \ce{^{137}Xe} of $10^{-2}\,\SI{}{kg^{-1}yr^{-1}}$ and $3\cdot 10^{-4}\,\SI{}{kg^{-1}yr^{-1}}$, respectively. These values are compatible with those shown in Tables \ref{tab:xe137} and \ref{tab:tritium}. However, unshielded xenon is exposed to the radiation coming from the fission and $(\upalpha,n)$ reactions from the rock and concrete walls, especially if it is placed near to them.   

We also recorded the production of several isotopes that could contribute significantly to the background of various science channels, specially at the low-energy electronic recoil region. For example, we noticed a non-negligible presence of nuclei from isotopic chains that start at \ce{^{121}Cs}, \ce{^{135}Sn} or \ce{^{123}Te}. More detailed simulations are ongoing to assess possible impact on the background of those isotopes in future studies.

\section{Summary and Conclusions}
\label{sec:conclusions}

We have performed Monte Carlo simulations of the cosmogenic background for the DARWIN experiment. The study was done for four underground laboratories that are candidates for the location of the detector.

We developed a custom-made DARWIN-Geant4 simulation framework in which we implemented new features to perform full muon simulations. First, we performed a detailed simulation of the underground experimental halls. Second, a muon generator that samples primaries using the realistic energy-angle correlation provided by the MUSIC-MUSUN software packages was developed.

We presented an estimate of the cosmogenic activation rates due to muon-induced neutrons in rock and concrete, \ce{^{137}Xe} and tritium. These isotopes are known to have non-negligible contributions to the background of physics channels such as the neutrinoless double beta decay and low-energy electronic recoils. In this work, we report a \ce{^{137}Xe} production rate $\sim 8$ times smaller than previous results for DARWIN \cite{DARWIN:2020jme}. This updated value sets the \ce{^{137}Xe} background at a level below the scattering of \ce{^{8}B} neutrinos with electrons at all the underground locations. Therefore, additional selection criteria have to be taken into account in the decision process leading to the final location of the DARWIN experiment. In addition, other xenon isotopes can be activated due to spallation processes, neutron captures from the muon secondary cascades or other inelastic processes; and they can also contribute to the background for processes such as the search of WIMPs. More detailed studies on the activation in the unshielded liquid xenon are being performed to determine their influence on the sensitivity of the detector to other physics channels of interest. 

\begin{acknowledgements}
This work was supported by the Swiss National Science Foundation under grants No 200020-162501 and No 200020-175863, by the European Union’s Horizon 2020 research and innovation programme under the Marie Sklodowska-Curie grant agreements No 674896, No 690575 and No 691164, by the European Research Council (ERC) grant agreements No 742789 (Xenoscope) and No 724320 (ULTIMATE), by the Max-Planck-Gesellschaft, by the Deutsche Forschungsgemeinschaft (DFG) under GRK-2149, by the US National Science Foundation (NSF) grants No 1719271 and No 1940209, by the PortugueseFCT, by the Dutch Science Council (NWO), by the Ministry of Education, Science and Technological Development of the Republic of Serbia and by grant ST/N000838/1 from Science and Technology Facilities Council (UK).
The authors would like to also thank Prof. V. Kudryavtsev for letting us use the MUSUN software.
\end{acknowledgements}

\bibliographystyle{JHEP}
\bibliography{mybibfile}

\appendix

\section{Detailed tables of the production rates}\label{app:tables}

In this work we have performed simulations with three different physics lists distributed with \geant{}: \texttt{Shielding}, \texttt{ShieldingLEND}, and \texttt{QGSP\!\_BIC\!\_HP}. Each list is constructed using hadronic and electromagnetic models together with a set of cross section databases. It is therefore possible to observe differences in the relative rates of processes such as neutron yields or isotope activation, reason why such differences have been propagated as uncertainties in these results.

\begin{figure}[h!]
	\centering
	\includegraphics[width=\linewidth]{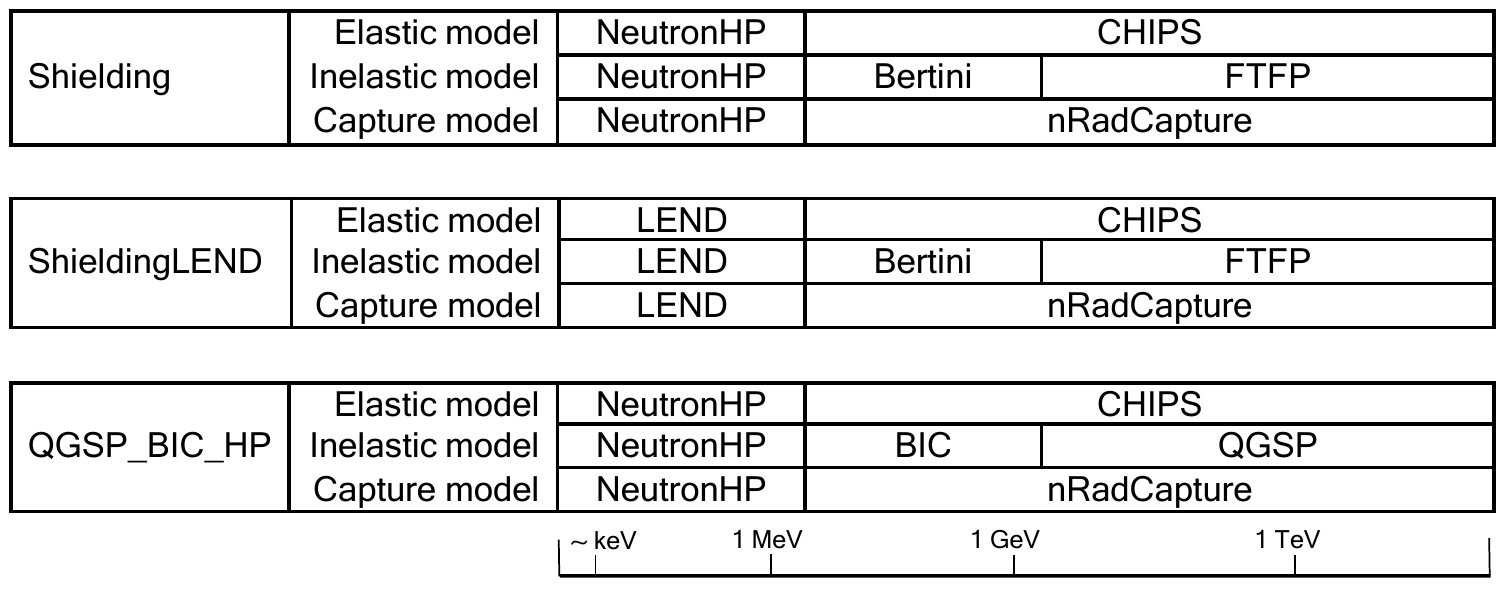}
	\caption{Hadronic models defined in the \geant{} physics lists considered in this study. The energy scale at the bottom is placed for visual reference. The detailed energy ranges for each model is found in the \geant{} physics list guide \cite{GeantPhys}. Cross sections libraries used for each model are not shown.} \label{fig:physlist}
\end{figure}

For reference, Fig. \ref{fig:physlist} shows a comparative table of the lists used in this work. For each list, the hadronic component consists of elastic, inelastic and capture-stopping models, depending on the energy of the particle. The name convention used on the table for the models is: Quark-gluon String with Precompound (QGSP), Fritiof Parton model with Precompound (FTFP), Binary Light Ion Cascade (BIC), CHiral Invariant Phase Space (CHIPS), Low Energy Nuclear Data (LEND), High Precision neutron model (HP).

\begin{table*}[!htbp]
\centering
\caption{Muon-induced \ce{^{137}Xe} production rate at the different underground laboratories given in $\SI{}{kg^{-1} yr^{-1}}$. We compare the results obtained with the \texttt{Shielding}, \texttt{ShieldingLEND} and \texttt{QGSP\!\_BIC\!\_HP} physics lists.} 
\begin{tabular}{lccc}
\hline
Site  & \texttt{Shielding} & \texttt{ShieldigLEND}  & \texttt{QGSP\!\_BIC\!\_HP}  \\
\hline
LNGS   & $8.22\cdot10^{-4}$ & $8.46\cdot10^{-4}$ & $6.87\cdot10^{-4}$  \\
SURF   & $1.42\cdot10^{-4}$ & $1.35\cdot10^{-4}$ & $1.13\cdot10^{-4}$  \\
LSM    & $1.65\cdot10^{-4}$ & $1.66\cdot10^{-4}$ & $1.23\cdot10^{-4}$ \\
SNOLAB & $6.75\cdot10^{-6}$ & $8.10\cdot10^{-6}$ & $6.75\cdot10^{-6}$  \\
\hline
\end{tabular}
\end{table*}

\begin{table*}[!htbp]
\centering
\caption{Muon-induced \ce{^{3}H} production rate at the different underground laboratories given in $\SI{}{kg^{-1} yr^{-1}}$. We compare the results obtained with the \texttt{Shielding}, \texttt{ShieldingLEND} and \texttt{QGSP\!\_BIC\!\_HP} physics lists.} 
\begin{tabular}{lccc}
\hline
Site  & \texttt{Shielding} & \texttt{ShieldigLEND}  & \texttt{QGSP\!\_BIC\!\_HP}  \\
\hline
LNGS & $1.16\cdot10^{-2}$ & $1.17\cdot10^{-2}$ & $1.33\cdot10^{-2}$  \\
SURF  & $1.82\cdot10^{-3}$ & $1.68\cdot10^{-3}$ & $2.45\cdot10^{-3}$  \\
LSM   & $2.17\cdot10^{-3}$ & $2.35\cdot10^{-3}$ & $2.79\cdot10^{-3}$  \\
SNOLAB    & $1.05\cdot10^{-4}$ & $1.63\cdot10^{-4}$ & $1.51\cdot10^{-4}$  \\
\hline
\end{tabular}
\end{table*}

\begin{table*}[!htbp]
\centering
\caption{Muon-induced activation rates (all processes) of \ce{^{125}Xe}, \ce{^{127}Xe}, \ce{^{133}Xe} and \ce{^{135}Xe} given in $\SI{}{kg^{-1} yr^{-1}}$. We compare the results obtained with the \texttt{Shielding}, \texttt{ShieldingLEND} and \texttt{QGSP\!\_BIC\!\_HP} physics lists.} 
\begin{tabular}{lcllll}
\hline
List  & Isotope & LNGS & SURF & LSM & SNOLAB \\
\hline
\texttt{Shielding}     & \ce{^{125}Xe} & $2.28\cdot 10^{-2}$ & $3.29\cdot 10^{-3}$ & $4.33\cdot 10^{-3}$ & $2.18\cdot 10^{-4}$ \\ 
\texttt{ShieldingLEND} &            & $2.41\cdot 10^{-2}$ & $3.36\cdot 10^{-3}$ & $4.94\cdot 10^{-3}$ & $2.59\cdot 10^{-4}$ \\
\texttt{QGSP\!\_BIC\!\_HP}   &            & $1.62\cdot 10^{-2}$ & $2.62\cdot 10^{-3}$ & $3.33\cdot 10^{-3}$ & $1.63\cdot 10^{-4}$ \\
\hline
\texttt{Shielding}     & \ce{^{127}Xe} & $6.39\cdot 10^{-2}$ & $1.02\cdot 10^{-2}$ & $1.25\cdot 10^{-2}$ & $6.38\cdot 10^{-4}$ \\ 
\texttt{ShieldingLEND} &            & $6.78\cdot 10^{-2}$ & $1.11\cdot 10^{-2}$ & $1.40\cdot 10^{-2}$ & $7.24\cdot 10^{-4}$ \\
\texttt{QGSP\!\_BIC\!\_HP}   &            & $5.59\cdot 10^{-2}$ & $9.52\cdot 10^{-3}$ & $1.18\cdot 10^{-2}$ & $6.01\cdot 10^{-4}$ \\
\hline
\texttt{Shielding}     & \ce{^{133}Xe} & $1.16\cdot 10^{-1}$ & $1.80\cdot 10^{-2}$ & $2.23\cdot 10^{-2}$ & $1.20\cdot 10^{-3}$ \\ 
\texttt{ShieldingLEND} &            & $1.20\cdot 10^{-1}$ & $1.82\cdot 10^{-2}$ & $2.45\cdot 10^{-2}$ & $1.29\cdot 10^{-3}$ \\
\texttt{QGSP\!\_BIC\!\_HP}   &            & $1.10\cdot 10^{-1}$ & $1.81\cdot 10^{-2}$ & $2.24\cdot 10^{-2}$ & $1.18\cdot 10^{-3}$ \\
\hline
\texttt{Shielding}  & \ce{^{135}Xe} & $4.63\cdot 10^{-2}$ & $7.42\cdot 10^{-3}$ & $9.10\cdot 10^{-3}$ & $5.01\cdot 10^{-4}$ \\ 
\texttt{ShieldingLEND} &            & $4.80\cdot 10^{-2}$ & $7.42\cdot 10^{-3}$ & $9.70\cdot 10^{-3}$ & $5.15\cdot 10^{-4}$ \\
\texttt{QGSP\!\_BIC\!\_HP}   &            & $4.47\cdot 10^{-2}$ & $7.82\cdot 10^{-3}$ & $9.34\cdot 10^{-3}$ & $5.17\cdot 10^{-4}$ \\
\hline
\end{tabular}
\end{table*}

\end{document}